# Microservices in IoT Security: Current Solutions, Research Challenges, and Future Directions

Maha Driss[a,b], Daniah Hasan[b], Wadii Boulila[a,b,*], Jawad Ahmad[c]

[a]*RIADI Laboratory, National School of Computer Sciences, University of Manouba, Tunisia*
[b]*IS Department, College of Computer Science and Engineering, Taibah University, Saudi Arabia*
[c]*School of Computing, Edinburgh Napier University, United Kingdom*

---

**Abstract**

In recent years, the Internet of Things (IoT) technology has led to the emergence of multiple smart applications in different vital sectors including healthcare, education, agriculture, energy management, etc. IoT aims to interconnect several intelligent devices over the Internet such as sensors, monitoring systems, and smart appliances to control, store, exchange, and analyze collected data. The main issue in IoT environments is that they can present potential vulnerabilities to be illegally accessed by malicious users, which threatens the safety and privacy of gathered data. To face this problem, several recent works have been conducted using microservices-based architecture to minimize the security threats and attacks related to IoT data. By employing microservices, these works offer extensible, reusable, and reconfigurable security features. In this paper, we aim to provide a survey about microservices-based approaches for securing IoT applications. This survey will help practitioners understand ongoing challenges and explore new and promising research opportunities in the IoT security field. To the best of our knowledge, this paper constitutes the first survey that investigates the use of microservices technology for securing IoT applications.

---

## 1. Introduction

The evolution of Internet technologies over the past years has led to the thriving of the Internet of Things (IoT). The IoT is defined by [1] as a "network of things, with clear element identification, embedded with software intelligence, sensors, and ubiquitous connectivity to the Internet". IoT applications allow physical objects called "Things" to connect from various smart devices in order to monitor, store, process, and analyze gathered data [2]. "Things" are computing devices that range from ordinary household objects to sophisticated industrial tools. Today, there are more than 8 billion connected IoT devices all over the world, and this number is expected to nearly triple in 2030 to more than 25 billion[*].

Nowadays, several domains and applications use the IoT for providing better services such as medical healthcare, business analytics, automotive industry, smart cities, smart agriculture, energy management, etc. [3] [4]. The IoT is playing a key role in these applications and domains by offering multiple solutions allowing to enhance people’s life. For instance, in the medical healthcare domain, IoT improves its quality while using added-value services that include: 1) remote medical consultations, 2) uninterrupted access to equipment, data, and patients’ information, 3) automatic transfer and analysis of data collected by devices, 4) and continuous monitoring of patients’ conditions, etc.

However, the benefits of IoT are constrained by many severe security issues and privacy threats [5] [6]. In fact, the interconnected devices within large IoT infrastructures are subject to several security attacks such as Distributed Denial of Service (DDoS) attacks, malware and ransomware, botnets, and phishing attacks. Therefore, implementing security and safety measures should be a top-notch priority for business environments that are based on IoT systems.

In the literature, several technical solutions have been proposed to deal with security issues in IoT environments. Among these solutions, we can list the adoption of the microservices-based architecture. Microservices technology is one of the leading technologies in the service computing field which is designed to expose the services in a decentralized and independent process. The microservices-based architecture allows different heterogeneous and distributed entities of an application to be developed, deployed, and scaled independently, thus becoming today a trend for IoT applications development. The integration of the microservices technologies within the IoT architecture enables enhancement of different aspects [7] [8]: 1) continuous delivery, testability, and deployability, 2) business functions reuse, 3) scalability and performance improvements, etc. Microservices-based applications can be designed to minimize their attack surface by performing only specific functions and performing them only when necessary so that fewer unused functions remain "active". Besides, Microservices can provide a higher level of isolation for IoT applications.

---

[*] https://www.statista.com/statistics/802690/worldwide-connected-devices-by-access-technology/

This paper aims to explore microservices capabilities employed to minimize the security issues in IoT environments by reviewing and comparing existing recent works.

The present paper is divided into four sections. In Section 2, an overview of important concepts related to IoT and microservices technologies is given. Section 3 provides a review of related research works relevant to the use of microservices-based architecture for securing IoT applications. Ongoing challenges and future research directions are presented in Section 4. Finally, Section 5 provides a summary of the present work.

## 2. Background

In the following subsections, we first present the IoT system definition, architecture, and security requirements. Next, we provide a detailed overview of the security attacks in IoT environments. Finally, we introduce the architecture and the essential characteristics of the microservices-based applications.

### *2.1. IoT system: definition, architecture, and security requirements*

IoT is an emerging paradigm that allows a group of physical nodes to exchange information over the network by using remotely connected objects or devices (e.g., smart lighting, smartphones, video cameras, and sensors, etc.). The IoT architecture, named also the IoT technology stack, involves four layers that are detailed as follows [2] [6]:

- ***Perception layer*** collects data from the environment using different distributed devices;
- ***Network layer*** ensures data routing and distribution over the Internet to various IoT hubs and devices by using internet gateways, cloud infrastructures, routing devices, etc;
- ***Middleware layer*** stores, examines, and processes data received from the network layer;
- ***Application layer*** is composed of a collection of problem-specific solutions that communicate with individuals, resolve issues, and interoperate with other remotely connected applications.

The IoT is a combination of different hardware and software technologies. Therefore, several fundamental characteristics emerge from the IoT paradigm [9] [10], including interconnectivity, support of dynamic changes, scalability, active engagement, intelligence, etc. In addition to the previously mentioned characteristics, security is a major requirement in any IoT ecosystem. IoT devices deal with numerous IoT users' data such as their locations, their contacts, their personal health records, their movements, and their purchasing preferences, etc. Besides, security threats of IoT devices are escalating as more devices are connected to the IoT networks. This fact leads to severe vulnerabilities and security threats with IoT environments [11] [12]. To improve IoT adoption, understanding security challenges and proposing effective solutions should be among the top-notch priorities in this field. In this context, several researchers have addressed the security requirements related to IoT systems based on the technologies that are employed as well as the categorized threats. In the literature, we distinguish two classes of security requirements identified for IoT systems [10] [12] [13]:

- ***Standard security requirements***: include confidentiality, access control, authentication, authorization, availability, key management, integrity, trust, accountability, and usability;
- ***Potential security requirements***: include scalability, privacy, identity, secure storage, secure content, manageability, decentralization, quality of service, reliability, mobility, and load balancing.

In essence, an IoT application should be responsible for satisfying all the security requirements needed in its ecosystem. Thus, security measurements should be developed in each layer of the IoT technology stack and also for their used software and hardware. These measurements aim to purposefully and effectively identify, protect, detect, respond, and recover from potential security threats and attacks.

### *2.2. Security attacks in IoT environments*

Security threats and attacks in IoT environments can be classified according to the different IoT architecture layers [10] [14]. Each layer could have its vulnerabilities that attackers typically target. In addition, since each layer depends on another layer, hence in case one layer is attacked, the other layers of the IoT stack could be affected. Therefore, security mechanisms should be implemented in each IoT layer to prevent potential attacks. Table 1 summarizes the most common security attacks that are prevalent in each IoT layer. These attacks are classified into four classes [15]:

- ***Physical Attacks (PA)***: can occur if the attacker is physically close to the IoT network or devices;
- ***Network Attacks (NA)***: occur by targeting the IoT network systems to cause damage;
- ***Software Attacks (SA)***: are performed by targeting the associated software or security vulnerabilities presented by an IoT application;
- ***Data Attacks (DA)***: target the computing resources allowing the maintenance of the connectivity between different IoT nodes and data collection that IoT resources and devices require to manipulate.

Table 1 shows that several vulnerabilities that exist in IoT environments can cause severe attacks ranging from attacks on IoT network devices to attacks on data exchanged between these devices. Each of these attacks could be remedied by applying the correspondent countermeasures [15]. Examples of countermeasures that are taken against data attacks could be the use of cryptographic access control and privacy-preserving authentication schemes. However, these countermeasures are personalized solutions that are strongly linked to the concerned IoT environment characteristics. Therefore, they could not present generic or reusable solutions to be applied in other different environments. To solve this issue, the microservices development paradigm is adopted since it allows to provide reconfigurable, reusable, lightweight, and scalable security features in heterogeneous and distributed environments like IoT environments.

Table 1. Common Security Attacks Related to Each Layer of the IoT Architecture.

| IoT Layer | Security Attack | Type of Attack | Brief Description | Possible Vulnerability |
|---|---|---|---|---|
| Perception Layer | Botnet attacks | PA + NA | Are computer programs that run automated and computerized commands over the Web | - Infrastructure vulnerabilities<br>- Human negligence<br>- Malware infection |
| | Malicious scripts | PA | Are fragments of script code targeting machines to make them vulnerable | - Operating system bugs<br>- Insecure devices |
| | Virus, Worms, and spyware | PA + SA | Known as malware which are malicious software attackers used to inject user's devices with the purpose of causing damage or stealing data | - Device hacking<br>- Insecure network<br>- Malware infection<br>- Human errors<br>- Insufficient security training |
| Network Layer | RFID spoofing | NA | Radio Frequency Identification uses the network waves to transfer data through the network. Attackers take advantage of the RFID signals to read, recode, and write data through the transmission process in the IoT network | - Data tracking<br>- Data corruption and deletion |
| | Routing information attacks | NA | Aim to block, replay, or spoof selected router messages to falsify their fields and contents | - Data alteration and corruption |
| **Middleware Layer** | Malicious node injections | PA | Attackers can inject malicious scripts which allows them to take control of the IoT nodes' operation process as well as the data flow between the nodes | - Operating system bugs<br>- Unpatched software<br>- Insecure devices |
| | Malicious code injections | SA + DA | Known as Remote Code Execution where an attacker exploits the software vulnerabilities by validating input flow and executing malicious code that targets specific applications. | |
| **Application Layer** | Data breaches | PA + DA | Sensitive or confidential information and data are stolen without authorization or user knowledge | - Human errors<br>- Poor security policies |
| | Shared technology vulnerabilities | PA + DA | Shared technology vulnerabilities cause several security issues including identification, authorization, access control, and availability that result in system weakness | - Insufficient device management<br>- Third-party vulnerabilities |
| | Cloud computing data security attacks | DA | Data stored in the cloud may be the target of unauthorized access | - Direct hacking<br>- Insecure cloud system |

### *2.3. Microservices: architecture and essential characteristics*

Over the last few years, the development paradigm of microservices has been used to describe an evolutionary process for developing software applications as packets of modular and independently deployable services [16] [17]. The primary purpose of adopting this paradigm is to break the business processes into small independent entities known as services [18]. Each service has its container, programming language, process, data storage, and communication mechanisms. This paradigm provides the software developers with guidelines and reduces the implementation efforts by offering a lightweight, flexible, and scalable process for building and running applications. Furthermore, the microservices architecture offers multiple other benefits [19] compared to monolithic software architectures [20] [21] such as the enhancement of service maintainability, deployability, testability, scalability, composability, and resilience. Thus, it enables organizations to quickly and efficiently develop more adaptable and reusable software solutions. Microservices technology started to gain popularity and multiple companies have adopted microservices-based architecture, such as Amazon, Netflix, Uber, etc., to enhance their business processes performance and gain agility and stability [22]. In this context, microservices' benefits have also appealed to IoT project managers and decision-makers. This is justified by the following reasons [23]:

- ***Flexibility and agile deployment***: microservice-based applications, which are employed to manage IoT networks and devices or to enable business logic functions, provide maximum flexibility and control over deployments. Since microservices are loosely coupled software entities, developers can decide to use only those they need, while deactivating the rest to save the IoT network computing resources;
- ***Resource-efficiency and portability***: containerized microservices are lightweight and can be scaled autonomously based on data workload, resulting in more efficient use of the IoT system resources. Additionally, microservices can be deployed on or off-premises, independently of each other, to better meet performance requirements and optimize the IoT system operations.
- ***Resilient operations and easy updates***: fine granularity, loose coupling, and containerization properties play a major role in mitigating the risks associated with executing IoT applications. Since the different microservices composing the IoT system operate separately from each other, the failure of a single microservice will not interrupt the normal operation of the entire system, making it highly resilient and secure to operate. Each microservice can also have its own and separate release cycle for easy maintenance and quick updates without requiring an immediate shutdown of the entire IoT system.

## 3. Related Works: Microservices-Based Solutions For Securing IoT Applications

The IoT is creating efficient, fast, and cost-effective applications that perform complex and value-added new functions. Due to its rapid evolution, the IoT faces several security issues that prevent its wide adoption. Recently, different technologies have been proposed and used to enhance the security of IoT applications. This section will review recent and relevant work on the use of microservices technology to secure IoT environments in order to understand their key capabilities and limitations.

In [22], the authors propose a high-level and abstract approach for developing general microservices-based IoT systems. To build secure, flexible, and extremely interoperable IoT systems, the proposed approach combines microservices' patterns, Application Programming Interface (API) gateways, distribution of services, and access control policies. To implement an Attribute-based Access Control (ABAC) model for the IoT systems, an Attribute-Based Encryption (ABE) scheme was proposed, which relies on a dynamic generation and implementation of multi-dimensional attributes-based policies. This scheme is also enforced by deploying identity management and authentication microservices as well as policy controller microservices.

In [24], a general IoT framework, which is based on a microservices architecture is proposed. This framework can easily extend, scale, and integrate third-party applications by using message-based and discovery-based mechanisms. To address security issues, the framework incorporates security microservices, which implement: role management, authentication and authorization, access control, and identity governance. They also offer features that ensure security monitoring, reporting, and auditing.

The authors in [25] propose a machine learning-based approach for modeling IoT service behavior that ensures the monitoring and the analysis of inter-service communication. This approach identifies the unusual services' behavior by employing microservices-based models. The role of these models is to observe the communication packets and intercept the malicious traffic that may result in security or safety risks. This approach uses two types of clustering algorithms, which are grid-based algorithms and k-means. The combination of microservices models, machine learning algorithms, and firewalling enables the implementation of access control in non-secure IoT installations.

In [26], the authors propose a graph-based access control model allowing to classify the communication traffic between microservices composing an IoT system as normal or anomalous. To implement this model, a traffic monitor service runs on each IoT node, active routers, and switches in the network. The role of the monitor service is to intercept each inter-service communication and perform Deep Packet Inspection (DPI). The DPI allows identifying the communication partners of each data transmission. Traffic monitor services are also enhanced by a firewalling mechanism to enable dropping the not permitted communication packets.

A smart surveillance system based on microservices architecture and blockchain technology is proposed by [27]. This system is an aggregation of a set of microservices offering different surveillance functions allowing object detection, tracking, and features extraction. These services are deployed on edge devices and managed by local fog nodes. Blockchain is adopted to provide a decentralized security mechanism allowing to secure the data exchanged among microservices. Smart Contracts are also used to record and attribute the authorization to access the monitoring data provided by microservices. Besides, the communication channels between edge nodes and fog nodes are protected using an encryption algorithm, which is the Advanced Encryption Standard- Rivest–Shamir–Adleman (AES-RSA).

In [28], a microservices-based architecture using blockchain technology is proposed to secure data access control in smart public safety. Security solutions are implemented as decentralized applications into separate containerized microservices that are built using Smart Contracts. These microservices are deployed on edge and fog computing nodes. Besides, blockchain-enabled security services are developed to: 1) execute consensus algorithms to verify transactions and generate new blocks and 2) run on single or multiple host machines to fulfill mining tasks independently.

To secure private IoT networks, [29] proposes the design and implementation of an edge-computing management solution, which is based on secure microservices. This solution relies on the edge gateway that is responsible for the secure management of devices, data, users, and configurations through microservices. The edge gateway incorporates API gateways and a session mechanism. The API gateway is used to filter all incoming requests to secure microservices. Besides, session-based authentication is implemented to secure the client support gateway.

Table 2. Summary of Related Works.

| Work | Case Study | IoT Security Requirement | IoT Component | IoT Layer | Security Solution |
|---|---|---|---|---|---|
| **[22]** | Personal health management + Autonomous vehicles | Access control | Devices and network | Middleware layer + Network layer | ABE scheme + multi-dimensional attributes-based policies + identity management, authentication, and policy controller microservices |
| **[24]** | - (A general framework is proposed) | Access control + Security monitoring | Devices | Network layer | Security microservices |
| **[25]** | Smart home | Communication data protection and security | Devices and network | Middleware layer + Network layer + Perception layer | Microservices-based models + Machine learning algorithms + Firewalls |
| **[26]** | Smart home | Access control + Communication data protection and security | Devices and network | Middleware layer + Network layer | Traffic monitor microservices + graph-based access control model + firewalls |
| **[27]** | Smart surveillance | Access control + Communication data protection and security | Communication nodes and network | Middleware layer | Microservices offering data encryption + Blockchain + Smart Contracts |
| **[28]** | Smart public safety | Data access control | Communication nodes | Middleware layer | Security microservices + Smart Contracts + Microservices-enabled private |
| | | | | | blockchain |
| **[29]** | - (A general solution is proposed) | Edge computing management security | Devices and network | Middleware layer + Network layer | Microservices + Edge gateway |

According to the summary provided in Table 2, most of the existing works have used microservices to fulfill a limited number of security requirements, which are mainly access control and communication data protection. Other essential requirements such as data integrity and quality of service are neglected. However, in an IoT network, it is important to implement security countermeasures that preserve its scalability and reliability to provide users with the best and effective services. Also, we notice that the proposed security solutions provided in the related works are implemented mainly in the middleware and network layers. Although, it is crucial to develop security measurements in each layer of the IoT technology stack to effectively protect and recover from potential security threats and attacks. Moreover, some related works like [24] and [29] are content simply with providing general solutions without conducting experiments on real case studies. These solutions could be ignored and not adopted in future research and industrial works since their effectiveness has not been proven. Based on the reasons mentioned earlier, it is evident that more in-depth investigations should be conducted to further explore the potential and promising capabilities of microservices technology to secure IoT applications. To this end, in the following section, we highlight the ongoing security challenges that still face the IoT environments and we propose multiple new lines of research to be explored to strengthen the security solutions implemented using microservices technology.

## 4. IoT Security Challenges and Future Research Directions

### *4.1. Ongoing challenges*

While the convergence of the microservices paradigm with IoT brings countless benefits, it also faces several challenges that need to be overcome to increase the adoption of microservices architectural style in various IoT applications. This section presents the key ongoing challenges that prevent the adoption of microservices with IoT, as detailed in the following points:

- ***Heterogeneous distribution and interoperability***: microservices-based applications are composed of fine-grained, distributed, and independent entities. These entities are implemented using different technologies, their data are stored in different databases, and communicate with each other through APIs that are independent of machine architecture and even programming language, which increases the total number of possible critical points that could be subject to security attacks [30].
- ***Container's vulnerabilities***: containers are a simplified way to develop, test, deploy, and orchestrate microservices across various environments. However, they become a tempting target for hackers since they present multiple vulnerabilities [31]. The traffic resulting from the communication needed for containers' management and orchestration contains data that could be valuable for hackers. Besides, most of the major components forming a containerized environment (i.e., hosts, registries, and images) can be potential targets for hackers. A hacked container can be the gateway to other parts of the IoT system, including the system hosts, the system's remote devices, and other interoperable containers, etc.
- ***Cloud security threats***: cloud-native computing is the development approach adopted by IoT systems since it permits to build, run, and deploy their distributed and scalable entities. However, this approach introduces multiple security threats [32]. With so much data entering the cloud, and especially public cloud services, these data may be compromised or used inappropriately by hackers. Data breaches can cause serious damage that can affect the reputation and the trust of organizations. Also, they can potentially impact the short-term and long-term revenues as well as result in loss of intellectual property and significant legal liabilities.
- ***Authentication and authorization***: Microservices-based applications should be able to verify the authenticity of each service involved in the communication because if one service is controlled by a hacker, the remaining services can be affected and can be misused maliciously [33]. In addition, when a service receives a message, it needs to know whether the message is suspicious or not and whether the source service which issued the message has valid authority. To address these issues, a set of authorization and access control mechanisms (e.g., policies, models, and protocols) should be implemented in order to fulfill the security requirements needed for heterogeneous environments (e.g., multi-cloud environments) and to ensure centralized security management that allows the deployed applications to be protected effectively.

- ***Resource limitations***: Most IoT devices are low-energy embedded devices that lack the computing resources needed to support the implementation of advanced and effective authentication and encryption algorithms because they are unable to perform complex processing operations in real-time [6]. To solve this issue, dedicated security chips can be used, but it increases cost, complexity, and energy consumption, so manufacturers have no incentive to go this route, especially in consumer-grade devices. Local storage is also limited and IoT devices often rely on cloud storage to store the data they generate; this fact opens the door for additional threats in the form of cloud security threats.

### *4.2. Future research directions*

This subsection provides a list of future research directions, which could be more investigated to enhance security mechanisms for IoT environments:

- ***Blockchain as a service***: To overcome the above challenges, blockchain technology can operate as a third party to secure the distributed and heterogeneous ecosystems of microservices-based IoT applications. Blockchain has proved its efficiency in this field by offering several features (e.g., tamper-proof, robust level of encryption, transparency, fast processing of transactions and coordination among billions of connected devices, anonymity, etc.) and benefits (e.g., improved trust and security, susceptibility to manipulation, and fairness, etc.). However, the traditional blockchain integration approach presents multiple disadvantages (e.g., scalability, energy wastage, and communication overhead and synchronization, etc.). To address these issues, Blockchain can be used as a Service (BaaS) by IoT applications in the same way that cloud computing is offering Software-as-a-Service (SaaS), Platform as a Service (PaaS), and Infrastructure as a service (IaaS) [34]. Numerous features can be created, hosted, and utilized dynamically in IoT contexts such as smart contracts, cryptography algorithms, identity access management platforms, identity-based consensus mechanisms, etc. BaaS features can be offered by different cloud providers and can be integrated into IoT devices as PaaS or SaaS.
- ***AI/ML-based solutions***: In the IoT context, Artificial Intelligence (AI) and Machine Learning (ML) have proved their efficiency in analyzing data that are gathered and stored at cloud infrastructures to provide rapid and accurate predictions and decisions. Besides, AI and ML could be adopted to improve the security in IoT ecosystems by allowing them to identify, investigate, and predict potential external threats [35]. By providing decentralized learning systems, AI/ML-based solutions can enhance the security in microservices-based IoT applications by monitoring the flow of data needed to ensure the communication and orchestration between these entities' containers.
- ***Trust management***: In the IoT applications context, traditional cloud systems require third-party service providers to build trust between data owners and users. To address this issue of trust, a proxy re-encryption scheme incorporating dynamic Smart Contracts at runtime can be designed and implemented [36]. Smart Contracts technology is defined as a set of protocols that digitally ensure the verification, control, and execution of agreements. Smart Contracts offer multiple benefits in terms of autonomy, safety, and accuracy and thus they are widely adopted in a variety of vital sectors such as healthcare, financial services, and supply chain, etc. In the context of IoT applications, the integration of smart contracts will eliminate the need for a trusted third party. Also, it effectively ensures the visibility of data between an owner and the user registered with the smart contract by using the proxy re-encryption. Thus, the proposed system allows secure sharing and storage of IoT sensor data. This is achieved by encrypting the data before uploading it to the cloud and re-encrypting it before sharing it with a user, thereby gaining total confidentiality.
- ***Ethical design for IoT***: Ethics is a philosophical branch that differentiates between what is morally right or wrong, just or unjust, while rationally justifying our moral judgments. Due to the complexity, heterogeneity, and large scale of IoT systems, new ideas and reflections must be presented to define the appropriate regulations and policies ensuring effective data security and privacy in the context of this complex environment. Thereby, to secure IoT ecosystems, the creation of ethical frameworks is necessary to help understand what is appropriate and inappropriate and what is good and bad [10]. These frameworks will encompass mechanisms that must be relevant to heterogeneous ecosystems that include humans, autonomous and self-determining systems, virtual and physical devices and environments, etc.

- ***Security-by-design approach***: Security by design is an approach to software and hardware development where security is incorporated from the beginning of the initial development process, and not as a late addition accomplished after an incident caused by a security breach. In the context of IoT applications, adopting concrete and best practices, following exploitable development guidelines, and mapping security requirements to the lifecycle of IoT components present key considerations to implement security by design approach [37]. This will ensure that IoT components remain reliable and operational from design to end of life. This promising approach needs to be investigated more and particularly by exploring and implementing the security patterns that are proposed for microservices-based architectures.

## 5. Conclusion

In recent years, companies, governments, and users have increasingly relied on smart technology, which has attracted researchers' attention to develop IoT technologies and make them safe to use. The rapid emergence of IoT and its wide adoption in multiple vital fields are justified by its ability to build effective and economic enterprise applications that perform new and complex functionalities more efficiently. In this paper, we aim to provide a review of current security challenges that still threaten IoT environments. This paper focuses on reviewing works proposing the microservices-based development paradigm as a solution to minimize some of the ongoing security challenges. It demonstrates via different case studies presented in the recent literature how microservices can improve security in such heterogeneous and distributed environments. While the reported work has shown a great potential of microservices technology, there is still a long way to go towards a decentralized and lightweight comprehensive security solution for IoT applications. Thus, several promising and effective new ideas are discussed as research directions at the end of this paper to be explored and to be coupled with the microservices' paradigm to enhance the security in IoT environments.